\documentclass[useAMS,usenatbib]{mn2e}
\usepackage{epsfig}
\def\Pm{\mbox{\rm P}_M}
\def\Rm{\mbox{\rm R}_M}
\def\Rmc{R_{\rm crit}}
\def\Rey{\mbox{\rm Re}}

\newcommand{\be}{\begin{equation}}
\newcommand{\ee}{\end{equation}}

\title[Fluctuation dynamos and their Faraday rotation signatures]
{Fluctuation dynamos and their Faraday rotation signatures}
\author[Pallavi Bhat and Kandaswamy Subramanian]{Pallavi Bhat\thanks{E-mail:
palvi@iucaa.ernet.in} and Kandaswamy Subramanian\thanks{kandu@iucaa.ernet.in}\\
IUCAA, Post Bag 4, Ganeshkhind, Pune 411007, India.}
\begin{document}


\pagerange{\pageref{firstpage}--\pageref{lastpage}} \pubyear{2012}

\maketitle

\label{firstpage}

\begin{abstract}
Turbulence is ubiquitous in many astrophysical systems 
like galaxies, galaxy clusters and possibly even the filaments in the 
intergalactic medium. We study fluctuation dynamo action in 
turbulent systems focusing on one observational signature; 
the random Faraday rotation measure (RM) from radio emission of background sources
seen through the intermittent magnetic field generated by such a dynamo. 
We simulate the fluctuation dynamo in periodic boxes up to resolutions 
of $512^3$, with varying fluid and magnetic Reynolds numbers, 
and measure the resulting random RMs. 
We show that, even though the magnetic field generated is intermittent,
it still 
allows for contributions to the RM to be significant. 
When the dynamo saturates,
the rms value of RM is of order 40-50\% of the value expected 
in a model where fields of strength $B_{rms}$  
uniformly fill cells of the largest 
turbulent eddy but are randomly oriented from one cell to another. 
This level of RM dispersion obtains 
across different values of magnetic Reynolds number
and Prandtl number explored.
We also use the random RMs to probe the structure of the 
generated fields to distinguish the 
contribution from intense and diffuse field regions.
We find that the strong field regions (say with B $>$ 2$B_{rms}$) 
contribute only of order 15-20\% to the RM. Thus rare structures do not 
dominate the RM; rather the general 'sea' of volume filling 
fluctuating fields are the dominant contributors. 
We also show that the magnetic integral scale, $L_{int}$, which
is directly related to the RM dispersion, increases in all the runs,
as Lorentz forces become important to saturate the dynamo.
It appears that due to the ordering effect
of the Lorentz forces, $L_{int}$ of the saturated field 
tends to a modest fraction, $1/2-1/3$ of the integral scale
of the velocity field, for all our runs.
These results are then applied to discuss the 
Faraday rotation signatures of fluctuation dynamo generated fields 
in young galaxies, galaxy clusters and intergalactic filaments. 

\end{abstract}


\section{Introduction}

The plasma in disk galaxies and galaxy clusters are observed to be  
magnetised. Disk galaxies have a large scale component of the magnetic field 
ordered on kpc scales with a strength of several micro-Gauss ($\mu$G) 
and a somewhat larger random component with coherence scales of 
tens of parsecs \citep{Fletcher10,Beck12}. 
Statistical studies of Faraday rotation in several galaxy clusters
suggest that the intra cluster medium also hosts 
a random field, with coherence scales
of several kpc to ten kpc and a strength of several $\mu$G, 
which goes up to tens of $\mu$G at the center of cool core clusters
\citep{clarke_etal_01,murgia04,govoni_feretti04,
vogt_ensslin,Govoni_etal2010,bonafede10,kuchar_ensslin11}. 
Moreover, there is evidence
of ordered $\mu$G fields in high redshift galaxies at $z\sim1$;
inferred from the statistical excess of Faraday rotation seen 
in distant quasars which have a 
MgII absorption system in their spectra \citep{Bernet08}. 
Understanding the
origin of these ordered fields presents an important challenge.

Cosmic magnetic fields are thought to be generated by 
dynamo amplification of weak seed fields.
Dynamos convert the kinetic energy of fluid motions 
to magnetic energy. Dynamos are particularly easy to excite
in a sufficiently conducting plasma which hosts random or turbulent motions.
In galaxies, turbulence can be driven by randomly occurring 
supernovae
\citep{korpi_etal1999,avillez05,balsara05,gressel,
wu_etal09,gent12}.
In galaxy clusters 
and the general intergalactic medium, 
turbulence could arise from cluster mergers and structure formation shocks 
\citep{NB99,Ryu_etal2008,Xu_etal2009,Paul_etal2011,Iapichino_etal2011,vazza11,Ryu_etal2012}. 
Such cosmological simulations show that the resulting turbulent
velocities in the cluster plasma are highly subsonic and hence
nearly incompressible.
Evidence for cluster turbulence to be nearly incompressible also
comes from observations of pressure fluctuations \citep{schuecker04,
churazov12}, and upper limits based on the width of X-ray emission
lines \citep{sanders_et10,sanders_etal11,sanders_fabian12}.

Such 
vortical 
turbulent motions generically lead 
to what is referred to as a fluctuation or
small scale dynamo under modest conditions; that
the magnetic Reynolds number $\Rm$ exceeds a critical value $\Rmc$ of order
a few tens \citep{kaz,KA92,S99,Cho_vish00,HBD04,Schek04,BS05,Cho_etal2009,TCB11,BSS12}.
The fluctuation dynamo amplifies magnetic fields on the fast eddy 
turn over time-scales (typically much smaller than the age
of the system), on coherence scales smaller than the outer scale
of the turbulence. 
On the other hand, mean field or large-scale dynamos, 
which amplify fields 
correlated on scales larger
than the turbulent eddy scales, 
typically require more special conditions
(like turbulence to be helical),
and operate on a much longer time scale.
Thus the fluctuation dynamos will be important 
in all astrophysical systems, from young galaxies (where they
probably generate the first fields) to galaxy clusters and intergalactic
filaments (where conditions for mean-field dynamo action are likely to be 
absent).

The rapid amplification by fluctuation dynamos comes at a cost.
The field is squeezed
into smaller and smaller volumes, as rapidly as it is amplified,
and gets highly intermittent in the kinematic stage \citep{ZRS}.
A critical issue for astrophysical applications is how
coherent are the fields when the fluctuation dynamo 
saturates \citep{S99,HBD03,HBD04,Schek04,SSH06,Ensslin_vogt2006,CR09}. 
Using simulations done with large magnetic Prandtl numbers 
($\Pm=\Rm/\Rey\gg 1$), but small fluid Reynolds numbers ($\Rey$),
\cite{Schek04} argued that the fluctuation dynamo generated 
fields saturate with a folded structure,
where the fields reverse at the folds with the 
power concentrating on resistive scales
$l_d \sim l/\Rm^{1/2}$ ($l$ is the forcing scale of the turbulence). 
For large $\Rm\gg1$ typical of astrophysical
systems this would lead to negligible Faraday rotation measure (RM).
Simulations of \cite{HBD03,HBD04} (HBD) with $\Pm=1$ and a large 
$\Rm=\Rey = 960$,
found the magnetic correlation function 
$w(r) = \langle{\bf B}({\bf x})\cdot {\bf B}({\bf x+r})\rangle$ has a
correlation scale $\sim 1/6$ th of the scale of the corresponding 
velocity correlation function, but much larger than the resistive scale.
This seems consistent with a simple 
model of \cite{S99} (S99) for nonlinear saturation of small-scale dynamos, 
which predicts that the power in the saturated state concentrates
on scales $l_c \sim l/\Rmc^{1/2}$. One could then expect
significant RMs, as is also consistent with the results of
\cite{SSH06} (SSH) and \cite{CR09} (CR09).
The case when both $\Rey$ and $\Pm$ are large, as in 
galactic and cluster plasmas, is not easy to simulate.
\footnote{If one uses the Spitzer values for viscosity and resistivity,
one gets $\Rey\sim 10^7$, $\Rm \sim 10^{18}$ in the galactic interstellar 
medium, while $\Rey \sim 1$ and $\Rm \sim 10^{29}$ for cluster plasma 
\citep{BS05}. However, 
in galaxy clusters, the viscosity (hence $\Rey$) 
and perhaps the resistivity (hence $\Rm$) 
are likely to be set by plasma effects \citep{Scheko05} rather than coulomb 
collisions, and then $\Rey$ would increase and $\Rm$ would 
decrease, although their exact values are uncertain.}
Indeed, the saturation of fluctuation dynamos could
be quite different in large $\Rey$ turbulent systems which display what
is called `spontaneous stochasticity',  
compared to laminar high-$\Pm$ systems \citep{Eyink,Beres2012}.

Note that Faraday rotation measurements are crucial
to infer the presence of coherent magnetic fields.
Therefore it is especially
important to understand how much Faraday rotation is produced if
one sees a polarised radio source through the possibly intermittent 
magnetic field generated by a fluctuation dynamo? 
Addressing this question 
will form the focus of the present work.
Some work on the RM from fluctuation dynamos has been done
by SSH and CR09. We will
consider here higher resolution simulations (up to 512$^3$) 
compared to SSH. We however follow SSH in computing the RM by 
directly integrating along a large number of lines of sights 
(unlike CR09 who related the dispersion in RM to the energy spectrum assuming
isotropy). We also extensively examine the sensitivity of the RM obtained
from fluctuation dynamos to variation of both $\Rm$ and $\Pm$ (compared
to both SSH and CR09).
Moreover, unlike earlier woks, we will also resolve the contribution to 
the RM from high field structures (where the field is much
larger than the rms value) compared to the general volume 
filling field. This can also help probe 
the structure of the dynamo generated fields. 

\begin{table}
\setlength{\tabcolsep}{3.5pt}
\caption{The parameters for various simulation runs in dimensionless units. Here $k_f=1.5$ for all the runs. 
The $u_{rms}$ is that which obtains in the kinematic stage while $b_{rms}$ is
the average value in the saturated state.}
 \begin{tabular}{|c|c|c|c|c|c|c|c|}
\hline
\hline
Run & {\small Resolution} & $\eta\times10^4$ & $\nu\times10^4$ & $u_{rms}$ & $b_{rms}$ & $\Pm$ & $\Rm$ \\ 
\hline
A & $128^3$ & 4.0 & 4.0   &   0.13 & 0.044 & 1  &  208   \\
B & $256^3$ & 2.0 & 2.0   &   0.14 & 0.061 & 1  &  466   \\
C & $256^3$ & 2.0 & 10.0  &   0.14 & 0.061 & 5  &  466   \\
D & $256^3$ & 2.0 & 100.0 &   0.18 & 0.087 & 50 &  586   \\
E & $512^3$ & 2.0 & 2.0   &   0.13 & 0.054 & 1  &  426   \\
F & $512^3$ & 1.5 & 1.5   &   0.14 & 0.067 & 1  &  622   \\
G & $512^3$ & 1.5 & 15    &   0.15 & 0.080 & 10 &  675   \\
\hline
\hline
\label{xxx}
\end{tabular}
\end{table}

The next section presents the simulations that we have carried out
to use for the RM analysis.
Section 3 sets out the methodology for calculating 
the Faraday rotation measure from the simulations and the results 
are presented in section 4.
Application of these results to
astrophysical systems is considered in section 5.
The last section presents a
discussion of these results and our conclusions. 


\section{Simulations of Fluctuation Dynamos}

In order to study the Faraday rotation signatures of fluctuation dynamos,
we have run a suite of simulations using the 
\textsc{Pencil Code} (http://pencil-code.googlecode.com \citep{BD02,B03}).
The pencil code uses a sixth-order finite difference
in space and a third-order accurate time stepping method.
The continuity, Navier-Stokes and induction equations are solved
in a Cartesian box of a size $2\pi$ on a cubic grid with 
$N^3$ mesh points, adopting periodic boundary conditions.
The fluid is assumed to be isothermal, viscous, electrically conducting
and mildly compressible.
The code uses dimensionless quantities by measuring length in units of 
L$/2\pi$ (where L is the size of the box), speed in units of isothermal
sound speed $c_s$, density in units of initial value $\rho_0$, 
and magnetic field in units of $(4\pi \rho_0 c^2_s)^{1/2}$.
To generate turbulent flow, a random force is included manifestly 
in the momentum equation. In Fourier space, this driving force is transverse
to the wave vector ${\bf k}$ and localized in wave-number space about
a wave-number $k_f$. 
It drives vortical motions in a wavelength
range around $2\pi/k_f$, which will also be the
energy carrying scales of the turbulent flow. 
The direction of the wave vector
and and its phase are changed at every time step in the simulation
making the force almost $\delta$-correlated in time
(see \cite{HBD04} for details).
For all our simulations, we choose
to drive the motions between wave-numbers of 1 and 2, and thus
the average $k_f=1.5$.
This choice is motivated by the fact that we wish
to resolve the small magnetic field scale structures in any turbulent
cell as well as possible. 
The strength of the forcing is adjusted so that the rms Mach
number of the turbulence, $u_{rms}$ in the code (where velocity
is measured in units of the isothermal sound speed), is typically about 0.15.
This implies also that the motions are nearly incompressible.
The magnetic and fluid Reynolds number through out this paper are
defined by $\Rm= u_{rms}/\eta k_f$ and $\Rey = u_{rms}/\nu k_f$,
where $\eta$ and $\nu$ are the resistivity and viscosity of the fluid.
The magnetic Prandtl number is defined as $\Pm =\Rm/\Rey = \nu/\eta$.

\begin{figure}
\epsfig{file=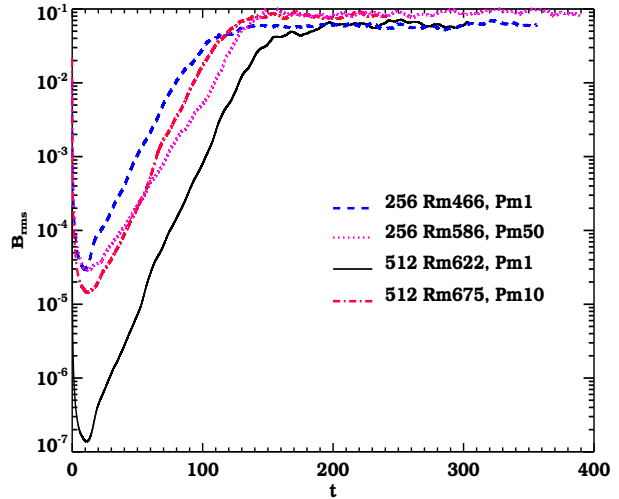, width=0.475\textwidth, height=0.3\textheight}
\caption{The evolution of $B_{rms}$ with time for fluctuation dynamo
simulations adopting $\Pm=1$ and $\Pm>1$ in both $256^3$ and $512^3$ resolutions.
The black solid line, blue dashed line,
red dash-dotted, and pink dotted lines represent 
simulation runs F, B, G and D respectively.}
\label{fig:growth}
\end{figure}

\begin{figure}
\epsfig{file=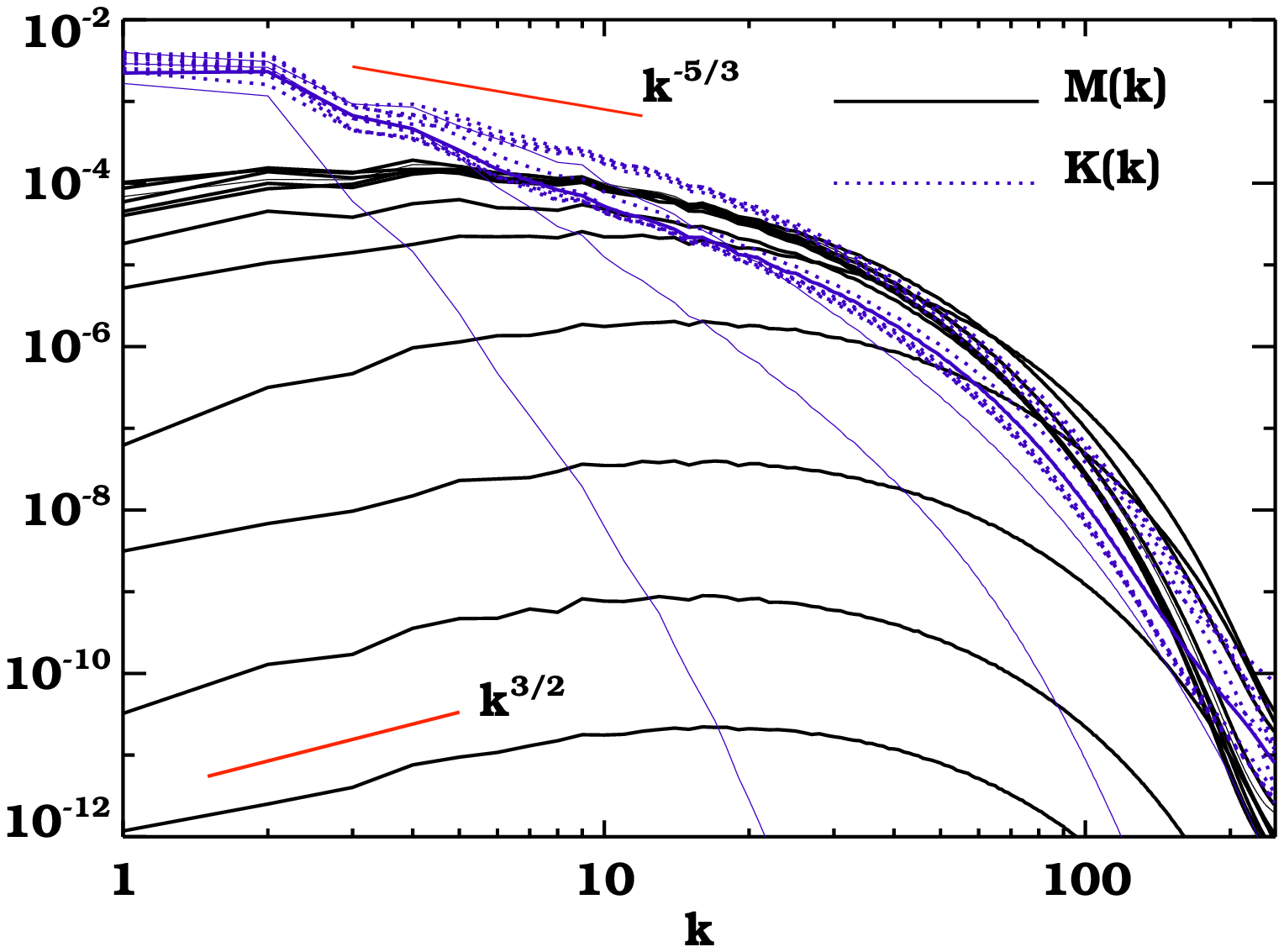, width=0.475
\textwidth, height=0.3\textheight}
\caption{The time evolution of the kinetic K(k) and magnetic M(k) spectra for
the $512^3$ simulation of fluctuation dynamos with $\Rm=\Rey = 622$ (run F). 
The first time for the spectra is $43 t_0$. The time difference between
successive spectra is about $\sim 22 t_0$. }
\label{fig:spectra}
\end{figure}

\begin{figure}
\epsfig{file=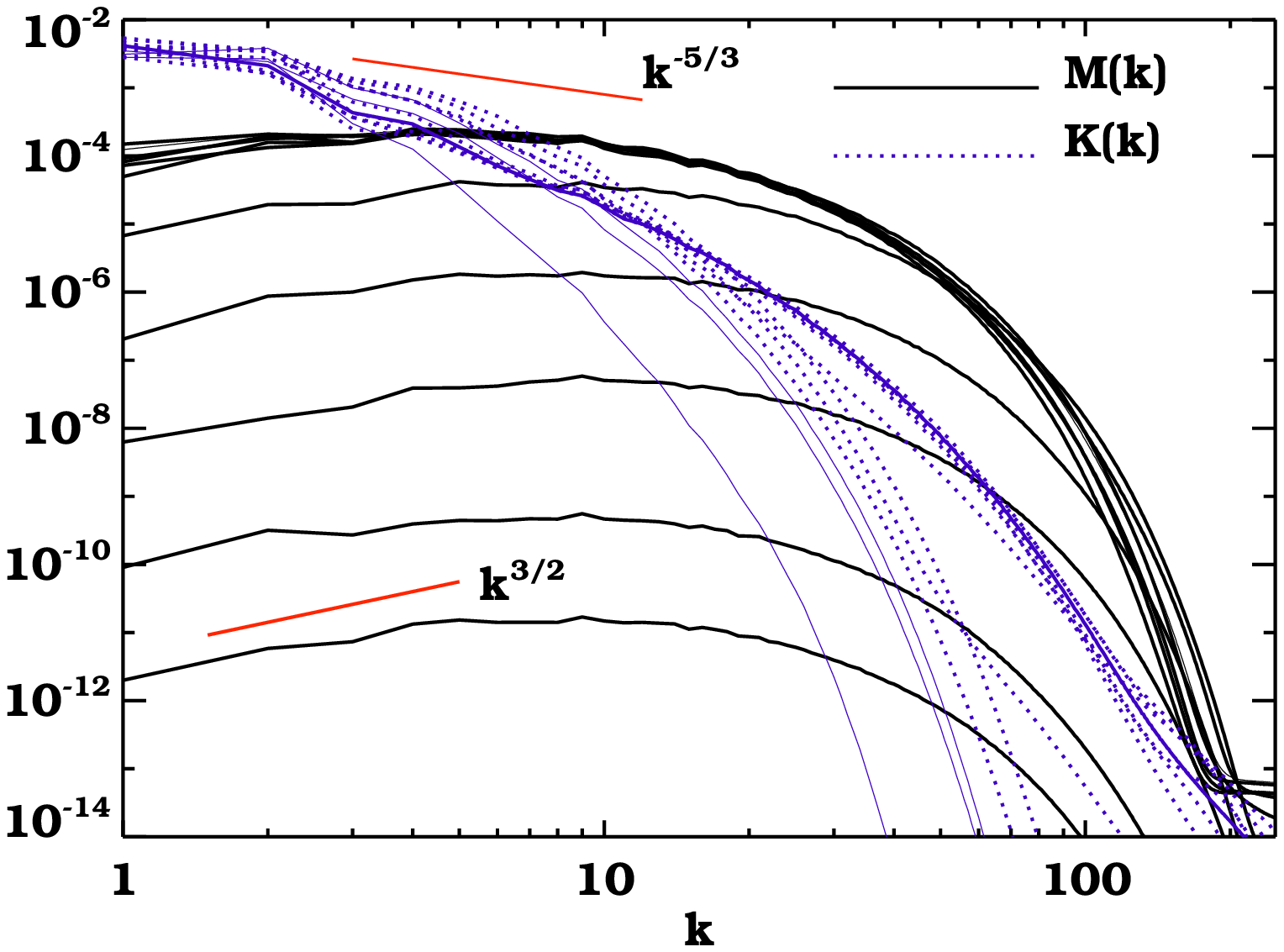, width=0.475
\textwidth, height=0.3\textheight}
\caption{The time evolution of the kinetic K(k) and magnetic spectra M(k) for
a $512^3$ simulation of fluctuation dynamos with $\Rm=675$, $\Rey=67.5$ 
(run G). 
The first time for the spectra is $45 t_0$. 
The time difference between
successive spectra is about $\sim$ 23 eddy turn over times}
\label{fig:spectra10}
\end{figure}

Starting with a weak Gaussian random seed magnetic field, 
and for $\Rm$ above a critical value,
the rms magnetic field $B_{rms}$, 
first grows exponentially
as shown in Fig.~\ref{fig:growth}, before saturating
(qualitatively similar to that by
HBD and in the cosmological context,
by \citet{beck_etal12}).
The time in this and other figures is measured in units
of the eddy turn over time $t_0 = (u_{rms} k_f)^{-1}$,
on the forcing scale $k_f$.
The simulation is allowed to run well 
into saturation as we want to calculate RM from the fields 
starting from the kinematic stage (when Lorentz forces are
not important) up to the saturated stage. 
We have run simulations with a resolution up to $512^3$ mesh points, 
with different $\Rm$ and $\Pm$ to be able to test the sensitivity of the 
resulting RM with respect to these parameters. 
These simulations adopt either $\Pm=1$ or $\Pm>1$
(Note that we have also considered $\Pm>1$ cases and not $\Pm<1$,
as the former case is more applicable to galactic and cluster plasmas).
We give in Table~\ref{xxx}, a summary of  
the parameters for all the runs. These include 
the number of mesh points $N$, $\eta$, $\nu$, the 
resulting $u_{rms}$ in the kinematic stage,
the average $b_{rms}$ at the saturated state, $\Pm$ and $\Rm$ calculated
using $u_{rms}$.

The time evolution of the kinetic and
magnetic spectra, $K(k,t)$ and $M(k,t)$ respectively, 
is shown in Fig.~\ref{fig:spectra}, for
one of our higher resolution ($512^3$) simulations, 
with $\Rm=\Rey=622$ (Run F). 
The magnetic spectra are shown as black solid lines while the kinetic
spectra as blue dotted lines, except for the final time,
where it is shown as a thick solid line.
(The build up of the kinetic spectra is also shown as 
thin solid lines for three early times.) 
The two short red solid lines with power
law behavior of the form $k^{3/2}$ and $k^{-5/3}$ are shown for comparison
with the Kazantsev and Kolmogorov spectra.
We see that the kinetic spectra eventually develops an
inertial range with a power law behavior slightly steeper than 
a Kolmogorov slope of $-5/3$. The magnetic spectra at early times 
have a Kazantsev form, $M(k) \propto k^{3/2}$ at small $k$, 
and are peaked at $k \sim 15$. However
as the field saturates the peak of $M(k)$ shifts to a much smaller $k \sim 4$,
with $M(k)$ decreasing with $k$ for larger $k$.
These spectra are qualitatively similar to those obtained
by HBD in their earlier work.

For $\Pm>1$ runs, we have kept $\eta$ the same as in the $\Pm=1$
runs of the corresponding resolution 
and increased $\nu$. The reason for not decreasing $\eta$ 
instead, is that the $\eta$ for various $\Pm=1$ runs are already 
set to almost their minimal values, if one takes care to resolve the 
smallest dissipative scales.
\footnote{
To deduce a reasonable estimate for $\Rm$ usable in the simulation 
for a given grid size, we argue as follows: 
Suppose we model $M(k)$ as a power law with $M(k) \propto k^s$.
Then, an estimate of the maximum value of $\Rm=R_{max}$ that can
be obtained in a simulation with N mesh points is 
$R_{max}\sim (k_f/k_{res})^{(s-1)/2}$, 
where $k_{res}=N/2$ assuming one needs to resolve
the dissipation scale where $\Rm(k)=1$ with at least 3 grid points.
For a Kolmogorov-like spectra, 
$s=-5/3$ and $512^3$ box, the estimated $R_{max}$ 
turns out to $\sim 950$.
In all our runs, we focus on being able to resolve smaller scales, 
and thus, conservatively do not exceed such estimates. 
Hence for $\Pm>1$ runs we increase 
the viscosity $\nu$, thus reducing the fluid Reynolds 
numbers from the case of $\Pm=1$.
}
The time evolution of the corresponding kinetic and
magnetic spectra, for a fluctuation dynamo simulation with $\Pm=10$
and $\Rm=675$ (run F), is shown in Fig.~\ref{fig:spectra10}.
The kinetic spectra cuts off much more sharply than $k^{-5/3}$, as 
the fluid is now much more viscous. 
The magnetic spectra are also flatter at early times 
than the Kazantsev form, although still peaked at a large $k\sim9$. 
As the field saturates the peak of $M(k)$ shifts again 
to a much smaller $k \sim 5$,
with $M(k)$ subsequently decreasing with $k$.
These spectra are also qualitatively similar to the high $\Pm$,
high $\Rey$ spectra presented in 
\cite{BS05}. We shall say more about these spectra later below.
We now turn to an analysis of these simulations to find the
RM predicted by the fluctuation dynamo.

\section{Faraday rotation measure from simulations}

The Faraday rotation measure (RM) is defined as  
\begin{equation}
{\rm RM} =  K \int_L n_e {\bf B} \cdot{\bf dl},
\label{frm}
\end{equation}
where $n_e$ is the thermal electron density,
${\bf B}$ is the magnetic field, 
the integration is along the line of sight `L' (LOS) from the source
to the observer, and $K= 0.81$ rad m$^{-2}$ cm$^{-3}$ 
$\mu$G$^{-1}$ pc$^{-1}$.
In our simulations, as the motions are nearly incompressible, 
the density is almost constant throughout the box 
\footnote{Note that as the turbulent scales are much smaller than
the system scale in general, $n_e$ can still vary on scales
larger than the scale of simulation box} (the rms density
fluctuations are of order a few percent),
and one can take $n_e$ out of the integral and denote it as $\bar{n}_e$.
We have checked that inclusion of density
in the integral to determine RM changes the result negligibly,
by less than 1$\%$. 

As in SSH, we directly compute, using the simulation data, 
$\int {\bf B} \cdot{\bf dl}$, and hence the RM over 3$N^2$ lines of sight,
along each $x$, $y$ and $z$-directions of the simulation box.
For example, if the line of sight integration is along $z$, at a given
location $(x_i,y_i)$, this involves
a discrete sum of $B_z$ of the form
\be
{\rm RM}(x_i,y_i,t) =  K \bar{n}_e 
\sum\limits_{j=0}^{N-1}  \left(\frac{2\pi}{N}\right)
B_z\left(x_i,y_i,\frac{2\pi j}{N},t\right). 
\label{frmz}
\ee
As the random magnetic field produced by the fluctuation dynamo is
expected to be nearly statistically isotropic, 
the mean value $\langle\int {\bf B} \cdot{\bf dl}\rangle$ over all the
lines of sight, and hence the mean RM is expected to nearly vanish. 
However the rms value of RM, which we denote as $\sigma_{RM}$ 
will be non-zero.

It is also convenient to normalise the RM by the rms value expected in
a simple model of the random magnetic fields.
For example, consider a model where a field of strength $B_{rms}$ 
fills each turbulent cell of scale $l=(2\pi/k_f)$ but is randomly oriented from
one turbulent cell to another. Also suppose that the LOS of length L, 
contains $M=L/l$ turbulent cells.
In such a model, 
we expect the mean RM to vanish but its dispersion to be given by 
\begin{eqnarray}
\sigma_{RM0} &=& K \bar{n}_e \ B_{rms} \ l \ \left(\sum\limits_{m,n=0}^{M-1}
 \langle\cos\theta_{m}\cos\theta_n\rangle \right)^{1/2} \nonumber \\
&=&  K\bar{n}_e 
\frac{B_{rms}}{\sqrt{3}} \
l\sqrt{\frac{L}{l}}
= K \bar{n}_e 
\frac{B_{rms}}{\sqrt{3}}
\left(\frac{2\pi}{k_f}\right) \sqrt{k_f}.
\label{frmo}
\end{eqnarray}
Here, $\theta_m$ is the angle between the LOS and the
magnetic field in each cell, labeled by the index $m$.
Also since $\theta_m$'s are independent and 
uniformly distributed over the solid angle  
we have $\langle\cos\theta_m\cos\theta_n\rangle = \delta_{mn}/3$.
Then only diagonal terms contribute to the sum, 
giving $\sum_{m,n}\langle\cos\theta_m\cos\theta_n\rangle = M/3 = (L/3l)$.
Moreover, for the last equality in Eq.~\ref{frmo}, 
we have replaced $l=2\pi/k_f$ and taken 
$L=2\pi$ the LOS length for the simulation box (see also SSH).
Thus the normalised RM defined as 
${\rm \overline{RM}} = {\rm RM}/\sigma_{RM0}$, is given by
\be
{\rm \overline{RM}}(x_i,y_i,t) =  
\sum\limits_{j=0}^{N-1} \left(\frac{2\pi}{N}\right) 
\frac{B_z\left(x_i,y_i, (2\pi j/N),t\right)}
{B_{rms}(t)(2\pi/k_f) \sqrt{k_f/3}}
\label{Nfrmz}
\ee
for a line of sight along the z-direction in the simulation box. 
This normalised RM is also expected to have a nearly zero mean, but a non
zero dispersion $\bar{\sigma}_{RM}$.
Due to the presence of $B_{rms}(t)$ in the denominator,
the normalised RM, $\bar{\sigma}_{RM}$, will not grow if 
$B_{rms}$ itself grows, but only
increase if the coherence scale of the field increases with time.

\begin{figure}
\epsfig{file=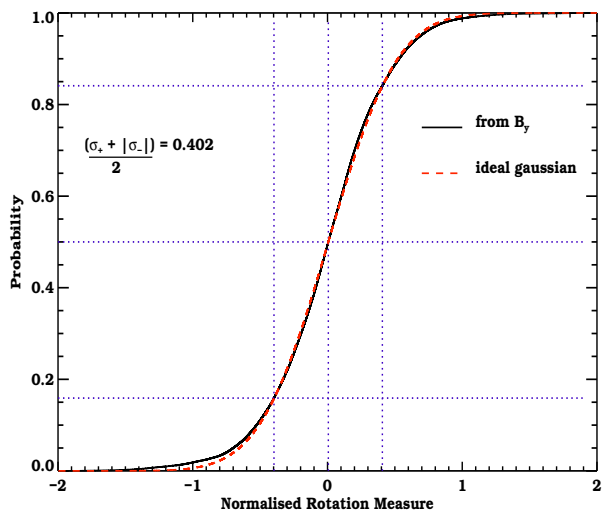, width=0.475\textwidth, height=0.3\textheight}
\caption{Cumulative probability distribution $C(x)$ of the normalised 
Faraday rotation measure $\overline{RM}$ from Run F at t=271$t_0$.
The horizontal lines show
probability of the mean $0.5$, and the one sigma levels 
of $0.159$ and $0.841$, assuming a Gaussian PDF. The vertical
values then indicate the corresponding mean $<\overline{RM}>$
and the $1\sigma_{\pm}$ values.
Average between $\vert\sigma_{\pm}\vert$ is derived as 0.402. 
The cumulative PDF of a Gaussian distribution with a mean $<\overline{RM}>$
and the averaged $1\sigma$ value, is also shown for comparison.} 
\label{cumpdf}
\end{figure}

In order to determine $\bar{\sigma}_{RM}$, we consider
the cumulative distribution of the $\overline{RM}$ for
the $N^2$ lines of sight in each direction.
Note that the cumulative distribution is preferred over 
the corresponding differential
probability distribution function (PDF) to avoid 
uncertainties which arise due to the choice of the bin size.
The cumulative probability distribution $C(X)$ 
can be determined by adding the number of 
occurrences of $\overline{RM} > X$, starting at the lowest
value in the data set, and 
normalising by the total number $N^2$ 
of data points (to convert to probabilities). 
We show in Fig.~\ref{cumpdf} such a
cumulative probability distribution $C(X)$ of  
$\overline{RM}$ for the $512^3$ simulation with $\Pm=1$ (run F), 
at 271$t_0$, after the dynamo has saturated.
We have chosen the lines of sight to be along the $y$-direction
of the simulation box.
The horizontal dotted lines show
probability of the mean $0.5$, and the one sigma levels 
of $0.159$ and $0.841$, assuming a Gaussian PDF. 
The $\overline{RM}$ values where these lines intersect the 
cumulative PDF, C(X) curve, give then values of
the corresponding mean $<\overline{RM}>$
and the dispersion, $\sigma_{-}$ and $\sigma_+$ respectively.
For a Gaussian PDF with zero mean, 
we expect $\sigma_{\pm}$ to be equal and opposite,
while for a general PDF their magnitudes can be different. 
We find the
average dispersion,  
$(\sigma_{+} +\vert\sigma_-\vert)/2$, and
then define $\bar\sigma_{RM}$ as its 
mean over all the 3 directions.
We can also obtain a normalised RM dispersion by constructing a
single cumulative PDF consisting of $\overline{RM}$s from all the 
3 directions. This matches closely with $\bar\sigma_{RM}$ defined above.
This is true for estimates made in both kinematic and saturated stages
(We will refer to this method of estimating $\bar\sigma_{RM}$
as method I). 
In the particular case shown in Fig.~\ref{cumpdf},
we have the $<\overline{RM}> = 0.006$, $\sigma_{+} = 0.401$,
$\sigma_- = 0.403$ and the average dispersion $=0.402$, obtained
from the magnetic field in y-direction.
The average dispersions calculated from fields in
x and z directions are 0.464 and 0.366 respectively, giving 
therefore, $\bar\sigma_{RM}$ = 0.411 (as can be seen from Fig.~\ref{RMc}).
The cumulative PDF of a Gaussian distribution, 
with the same mean $<\overline{RM}>$
and dispersion (averaged $1\sigma$ value), 
is also shown for comparison. We see in this case that 
$C(X)$ is quite well fit by the cumulative PDF of a Gaussian.
Note that the components of $B_i$ themselves are not expected to
have a Gaussian PDF \citep{BS05}, but the RM involves a sum
of $B_i$'s over a large number of mesh points. 
The PDF of this sum would then tend to a Gaussian if 
the $B_i$'s were independent or their correlation length were small
compared to the box size, due to the central limit theorem.
\footnote{
An interesting feature that we find is that when
only a part of the simulation box is considered for forming 
the cumulative PDF, its deviation from a Gaussian is large 
and $C(x)$ shows a large bias. One can infact get $\overline{RM}$
 larger than $\bar\sigma_{RM}$. This implies that the sampling of the 
entire turbulent cell is necessary to obtain a Gaussian 
cumulative PDF. Otherwise, when one is sampling only a part of 
the turbulent cell, correlated structures in the magnetic field 
can show up as a large bias in the PDF. Such a feature can be relevant
in interpreting the RM observations, as in \citet{murgia04}, where
the radio source could be extended over a scale smaller than the turbulent scale.
}

There are other cases when the Gaussian PDF does not provide 
a good fit to the wings of $C(x)$. Thus we also calculate for comparison
$\bar\sigma_{RM}$ directly as the standard deviation of the
set of $\overline{RM}(x_i,y_i,t)$ (henceforth method II).
A third method (method III) of estimating $\bar\sigma_{RM}$, 
which however assumes the
statistical isotropy of the random magnetic field generated by the
fluctuation dynamo, is to relate it to the integral scale of the field.
We have, using Eq. 9 of \citep{CR09} and Eq.~\ref{frmo} above,
\begin{equation}
\bar\sigma_{RM}=\frac{\sqrt{3}}{2}
\sqrt{\frac{L_{int} k_f}{2\pi}}
=\frac{\sqrt{3}}{2}
\sqrt{\frac{L_{int}}{l}},
\label{cr_frm}
\end{equation}
where $L_{int}$ is the integral scale  of the random magnetic field
and is defined by,
\begin{equation}
L_{int}(t) =\frac{\int (2\pi/k) M(k,t)dk }{\int M(k,t)dk}.
\label{lint}
\end{equation}

Note that the integral scale as defined here has the same order of
magnitude as the integral scales $L_L$ and $L_N$ 
defined respectively using the longitudinal and transverse correlation
functions. For any statistically homogeneous, isotropic, reflection
invariant and divergence free vector field,
$L_L= 2L_N = (3/8) L_{int}$ \citep{Monin_Yaglom}. 
Thus given the magnetic power spectra $M(k,t)$, one can
calculate the integral scale $L_{int}(t)$ and  
hence the normalised RM, $\bar\sigma_{RM}$.
One can also see that for a fixed $k_f$, the magnitude and 
evolution of $\bar\sigma_{RM}$ essentially reflects the 
evolution of the integral scale $L_{int}$.
This method is useful for estimating $\bar\sigma_{RM}$
from just the magnetic spectra, assuming 
statistical isotropy; however,
it cannot be used to separate the $\bar\sigma_{RM}$
contribution from high field structures versus the general
volume filling field (see below). 
We now turn to the results of the computations of $\bar\sigma_{RM}$
for the various simulations that we have performed, and their implications. 
 
\section{RM from fluctuation dynamos: results}

\begin{figure}
\epsfig{file=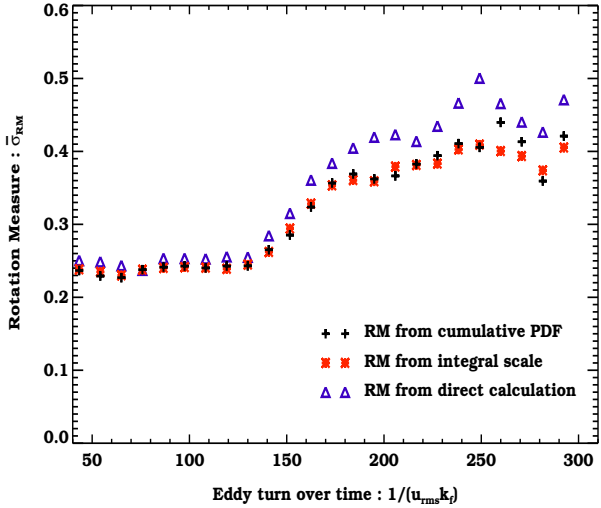, width=0.475
\textwidth, height=0.3\textheight}
\caption{The time evolution of the normalised RM ($\bar\sigma_{RM}$)
for the $512^3$ run (F), with $\Rm=\Rey=622$. The crosses show
the result of the direct calculation by shooting $3N^2$ lines of sight
through the simulation box. The triangles show the result of the direct estimate
of the standard deviation of $RM$, and the stars, the result of
integrating the energy spectrum (method III).}
\label{RMc}
\end{figure}

We begin by considering one of the runs of fluctuation dynamos 
with the highest value of $\Rm$, run F, with $512^3$ resolution
and $\Rey=\Rm=622$.
The time evolution of the normalised RM, $\bar\sigma_{RM}(t)$,
for this run is shown in Fig.~\ref{RMc},
starting from the kinematic stage to the saturation of 
the dynamo. The results are shown for all three methods of
calculating $\bar\sigma_{RM}$. The crosses show
the result of calculating RM by shooting $3N^2$ lines of sight
through the simulation box (method I), the triangles the direct estimate
of the standard deviation of $\overline{RM}$ (method II), 
and the stars the result of integrating the energy spectrum (method III).

First, we find that all three estimates of $\bar\sigma_{RM}$
agree reasonably well, a closer agreement being
obtained between methods I and III. The agreement between method I, which
uses a configuration space analysis and
method III using the Fourier space spectrum is reassuring.
The direct estimate
of the standard deviation (method II) 
also agrees with the other methods at early times, but later
as the dynamo saturates, 
always gives a larger
estimate of $\bar\sigma_{RM}$ by about $10\%-20\%$. 
This indicates that after saturation, there is usually
an excess of $\overline{RM}$ in the wings of the cumulative PDF 
over and above that predicted by a Gaussian approximation to $C(X)$.
This excess could arise due to the increase in the magnetic 
correlation scale and the deviation from statistical isotropy
when the dynamo saturates.

\begin{figure}
\epsfig{file=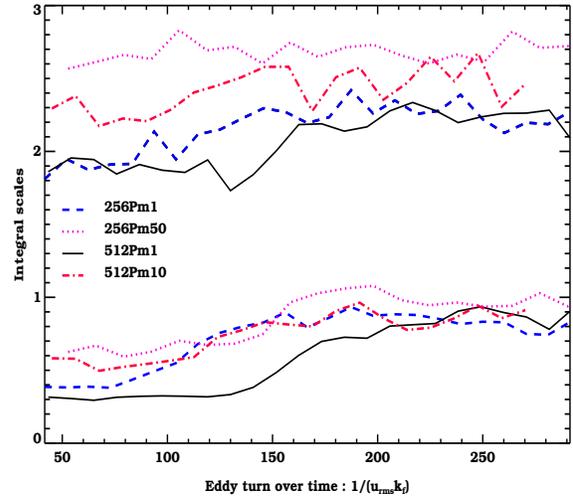, width=0.475\textwidth, height=0.3\textheight}
\caption{Comparison of integral scale for runs B, D, F and G. Lines on the 
upper half of the plot correspond to the velocity integral scales, $L^V_{int}$ and on the lower
half correspond to the magnetic integral scales, $L_{int}$. The linestyles are matched with
those in Fig.~\ref{fig:growth} to be able to compare the times at which
the integral scales start growing to the corresponding regime in the magnetic field growth.}
\label{intscale}
\end{figure}

We see from Fig.~\ref{RMc} that the normalised Faraday rotation
measure is almost constant in the kinematic stage with 
$\bar\sigma_{RM} \sim 0.24$. This obtains even though $B_{rms}$ 
itself is growing exponentially. It indicates that during the 
kinematic stage, the spectrum $M(k,t)$ evolves in a self-similar 
fashion, maintaining the integral scale. However by the
time the dynamo saturates, there is a substantial increase in the normalised RM to the value $\bar\sigma_{RM} \sim 0.4-0.5$. 
Since $\bar{\sigma}_{RM}^2$ is directly proportional to the 
integral scale $L_{int}$ (in method III), 
this implies that $L_{int}$ has increased by a factor
of $\sim 3$ as one goes from the kinematic to the
saturated state. To check this, we have shown in  Fig.~\ref{intscale},
the time evolution of both the magnetic and kinetic integral scales for various runs. We can see that for the $\Pm=1$, $\Rm=622$ run, 
$L_{int}$ does increase from a value of about $0.3$ in the kinematic stage
to $L_{int} \sim 0.9$, or a factor $\sim 3$, by the time the dynamo saturates
(see the thick solid line at the bottom of Fig.~\ref{intscale}).
In contrast, we find that the corresponding integral scale of the
velocity field $L_{int}^V$ (defined as in Eq.~\ref{lint} with $M(k)$ 
replaced by $K(k)$), only grows from about 2 to a value of $2.2$ during 
the same period (the black solid line in the upper half of 
Fig.~\ref{intscale}).

Importantly, a comparison of Fig.~\ref{RMc} and Fig.~\ref{intscale} 
with Fig.~\ref{fig:growth} shows that
the magnetic integral scale and the $\bar\sigma_{RM}$ begin to increase
at $t/t_0\sim 150$ just as the field begins to saturate due to
the influence of Lorentz forces. 
In fact a comparison of Fig.~\ref{intscale} and Fig.~\ref{fig:growth}
shows that for all the cases we have considered, the magnetic integral
scale increases when Lorentz forces become important, whereas the
integral scale of the velocity field changes very little during this
period. 
Thus, clearly, it is the influence
of the Lorentz forces that leads to larger and larger coherence
scale of the magnetic field reflected in the increase of $L_{int}(t)$
and $\bar\sigma_{RM}$. 

The value of $\bar\sigma_{RM} \sim 0.4-0.5$ that we
obtain is quite significant given that one expects the
fluctuation dynamo generated field to be fairly
intermittent. It implies that 
the rms value of RM in the saturated state of the fluctuation dynamo,
is of order 40\%-50\%, of that expected in a model
where $B_{rms}$ strength fields volume fill each turbulent
cell, but are randomly oriented from one cell to another.
We will apply this result in section~\ref{astro_app} 
to discuss the RM obtained
in various astrophysical systems.

\subsection{Sensitivity of RM to $\rm{\Rm}$ and $\rm{\Pm}$}

\begin{figure}
\epsfig{file=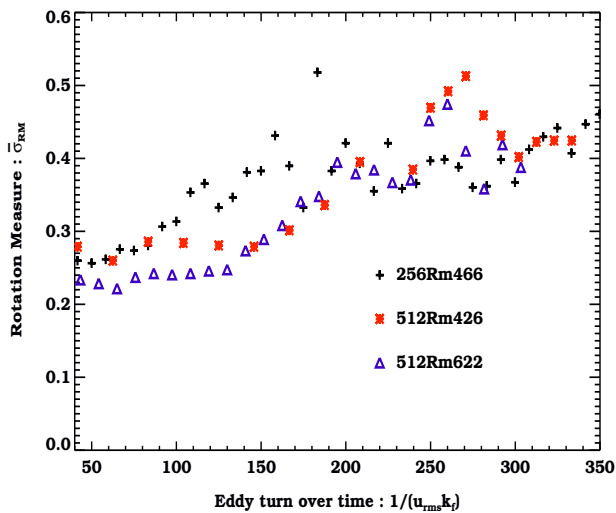, width=0.475\textwidth, height=0.3\textheight}
\caption{Sensitivity of RM to $\Rm$ : The time evolution of $\bar{\sigma}_{RM}$ calculated using method I is shown for
simulations (runs B, E, F) with different $\Rm$ keeping $\Pm=1$.
} 
\label{sensRM}
\end{figure}

\begin{figure}
\epsfig{file=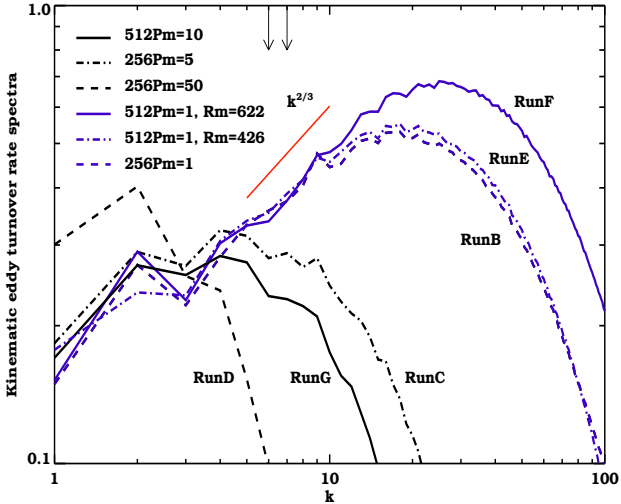, width=0.485\textwidth, height=0.30\textheight}
\caption{Comparison of eddy turn over rate spectra from different runs, in kinematic stage.
Spectra are plotted at $t\sim83t_0$ for Run B, $t\sim64t_0$ for Run C and 
$t\sim80t_0$ for Run D, time $t\sim43t_0$ for Run F, $t\sim45t_0$ for Run G.}
\label{eddy:spectra:kin}
\end{figure}

It is important to test the sensitivity of  
the RM produced by the fluctuation dynamo generated fields, 
to changes in the values of the magnetic Reynolds
number and Prandtl number. 
For testing the sensitivity of our results to $\Rm$, 
we have run another high resolution ($512^3$) 
simulation with a lower $\Rm=426$
and $\Pm=1$ (run E) to compare with run F, where $\Rm=622$. 
The results of RM analysis for both these simulations are compared
in Fig.~\ref{sensRM}, where we plot the time evolution of 
$\bar{\sigma}_{RM}$ obtained from
these runs. To check the effect of the resolution, the
figure also shows the results from a $256^3$ run having a 
similar $\Rm=466$ (run B). We see that $\bar{\sigma}_{RM}$ for the 
lower $\Rm$ runs start off with a higher value in the kinematic stage
as expected, if the integral scale is initially larger.
Such an expectation is consistent with what is seen in Fig.~\ref{intscale}
(compare the black solid and blue dashed lines).
The reason for this larger $L_{int}$ in the kinematic stage 
is probably due to the fact that only slightly larger scale eddies 
are able to amplify the field (these are the eddies
for which $\Rm(k)$ defined as $u_{rms}(k)/k\eta$ or $\sqrt{k K(k)}/k\eta$, is 
greater than $\Rmc$) in the lower $\Rm$ cases, compared to
the case when $\Rm=622$ (see also the discussion below of 
Fig.~\ref{eddy:spectra:kin}).

Again as the field begins to saturate, $\bar{\sigma}_{RM}$ increases
due to the ordering effect of Lorentz forces, and asymptotes to a value 
between $0.4-0.5$, for run B and E as well. It is of interest to note 
that for run B where the
Lorentz forces become important at an earlier time compared to run F and E, 
the rise of $\bar{\sigma}_{RM}$ also begins earlier. 
Our results are thus consistent with the
idea that $\bar{\sigma}_{RM}$ obtained in the saturated state is 
independent of $\Rm$, although we have explored at present only a 
modest range of $\Rm$. 

It is of interest to compare our results 
with that obtained from an independent 
$\Rm=1784$, $\Pm=1$, $1024^3$ 
simulation of the
fluctuation dynamo whose data is publicly available online
at the JHU turbulent database \citep{Li_etal2008, Perlman_etal}.
These authors simulate the fluctuation dynamo using a forcing
in the form of a Taylor-Green flow with $k_f=2$. They give the integral
scales of the velocity and magnetic fields in the saturated
state of the dynamo, to be $L_{int}^V = 1.5$ and 
$L_{int}=0.93$ respectively (in our definition). 
We can use this data in Eq.~\ref{cr_frm} 
to estimate the $\bar\sigma_{RM}$. We get 
$\bar\sigma_{RM} =0.47$, which is remarkably consistent
with our results above.
As this is an independent simulation with a much higher $\Rm$,
it would appear that a $\bar\sigma_{RM} \sim 0.4-0.5$ is a
robust result, atleast for the $\Pm=1$ case.


\begin{figure}
\epsfig{file=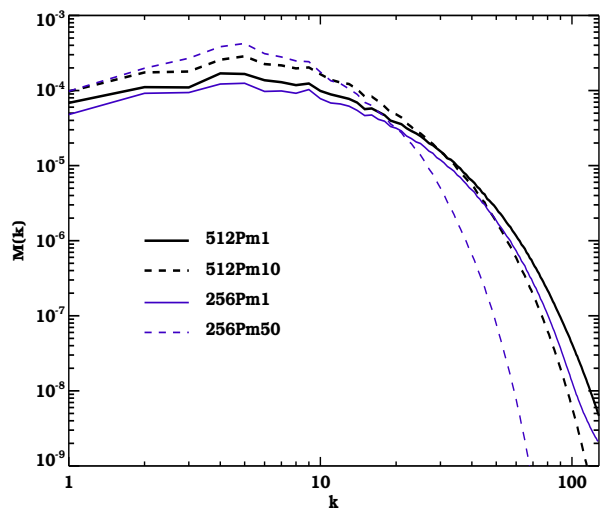, width=0.475\textwidth, height=0.3\textheight}
\caption{Comparison of the final saturated magnetic spectra from different $\Pm$ runs (B, D, E and F).}
\label{pm:spectra}
\end{figure}

\begin{figure}
\epsfig{file=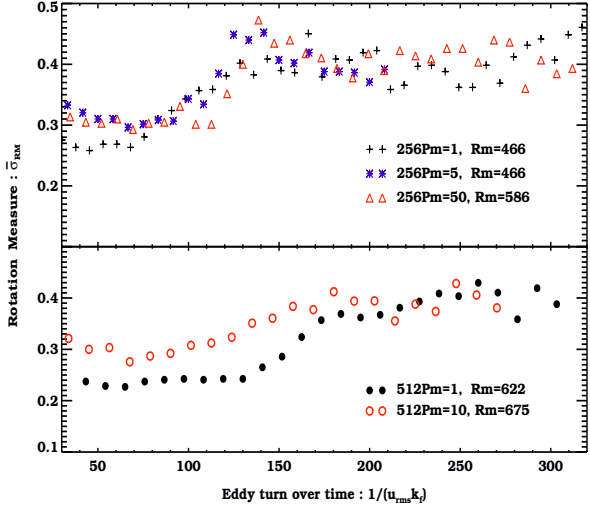, width=0.475\textwidth, height=0.3\textheight}
\caption{Sensitivity of RM to $\Pm$ : The top panel shows time evolution of $\bar{\sigma}_{RM}$ for
simulations with different $\Pm=1,5,50$, for runs (B, C and D) having the
same resolution, $256^3$. Bottom panel shows time evolution of $\bar{\sigma}_{RM}$ (using method I) for
simulations of higher resolution $512^3$, with $\Pm=1$ and $10$ (run F and G).
} 
\label{sensPM}
\end{figure}

In order to test the sensitivity of our results to having higher $\Pm$,
we have also run a suite of simulations with the same resolution ($256^3$),
but with varying $\Pm=1,5,50$ (runs B, C, D), and also one with a higher
resolution ($512^3$) with $\Pm=10$ (run G) (Note that in galactic and 
cluster plasmas, one expects $\Pm \gg 1$). In contrast to \cite{Schek04}, 
we have increased $\Pm$ by increasing the viscosity $\nu$ 
(decreasing $\Rey$), while keeping the resistivity constant across runs.
This is in keeping with the needed resolution at small scales, 
as explained in section 2 (We have nevertheless kept $\Rey \gg1$ for all
the runs).
Interestingly, when $\Rey$ is decreased keeping $\Rm$ almost the same
(so as to increase $\Pm$), 
we find that the magnetic spectra in the saturated state, 
seems to have lower energy at the small `resistive' scales 
for a larger $\Pm$; see Fig.~\ref{pm:spectra}. 

In the top panel of Fig.~\ref{sensPM}, we show the time evolution of 
$\bar{\sigma}_{RM}$ for the runs with $256^3$ resolution. 
We also
show the $\bar{\sigma}_{RM}$ evolution for the higher resolution 
runs with $\Pm=1$ and $\Pm=10$, separately in the bottom panel.
In the kinematic stage, we find that $\bar{\sigma}_{RM}$ tends to be larger
for higher $\Pm$, but also smaller, the higher the $\Rm$.
This again basically reflects
the corresponding dependence of the magnetic 
integral scale on these parameters (see Fig.~\ref{intscale}).
The integral scale of the velocity field itself 
(shown in Fig.~\ref{intscale}), 
is expected to be 
larger for the larger $\Pm$ case (assuming the same forcing scale),
since a larger viscosity (for the high $\Pm$ run) 
damps more of the small scale power in the velocity field.

In order to understand the reason for a larger magnetic integral scale 
in the kinematic stage for the higher $\Pm$ run, it is instructive 
to look also at Fig.~\ref{eddy:spectra:kin}.
Here we have given the spectra of eddy turn over rate
defined as $\gamma(k) = k\sqrt{kK(k)}$ at 
times when the dynamo
is still in the kinematic stage. We see that $\gamma(k)$ rises
with $k$ till about $k\sim 25$ for the run F with $\Pm=1$. While
for run G with $\Pm=10$, $\gamma(k)$ is maximum and 
relatively flat between $k \sim 2-5$. Note that eddies with a scale such that
$\gamma(k)$ is larger will tend to grow the field first, provided
their corresponding magnetic Reynolds number $\Rm(k)$ is super critical.
For run F, this happens for $k$ smaller than a critical value 
$k_{crit} = 7$, while for runs B, E and G, $k_{crit}=6$ 
(we have marked the $k_{crit}$ for these runs by arrows in 
Fig.~\ref{eddy:spectra:kin}).  
Since such eddies have a smaller $k$ for the $\Pm=10$ run compared
to the $\Pm=1$ case, we do expect a larger $L_{int}$ for   
former case compared to the latter, during the kinematic evolution.
\begin{table}
\setlength{\tabcolsep}{3.5pt}
\caption{Summary of the results obtained from various simulation runs. The individual integral
scale values are obtained by averaging the results over each stage 
(kinematic and saturated), where the evolution curves are relatively flat (i.e. the period
where these scales increase, is not considered for averaging). Similarly, the
$\bar{\sigma}_{RM}$ values are obtained by averaging the results over the saturated stage.
The independent JHU simulation ($1024^3$) result has also been shown 
in the last row for comparison. Note that the forcing scale $k_f$ = 2 for their run.
}
 \begin{tabular}{|c|c|c|c|c|c|c|c|c|c|}
\hline
\hline
& & & & \multicolumn{2}{|c|}{Kinematic} & \multicolumn{2}{|c|}{Saturation} & \multicolumn{2}{|c|}{$\bar{\sigma}_{RM}$} \\
\hline
Run & {\small Res} & $\Pm$ & $\Rm$ & $L_{int}$ & $L^V_{int}$ & $L_{int}$ & $L^V_{int}$ & PDF & Direct \\ 
\hline
A & $128^3$ &  1  &  208 & 1.0 & 2.8 & 1.2 & 2.8 & 0.41 & 0.50  \\
B & $256^3$ &  1  &  466 & 0.4 & 1.9 & 0.8 & 2.2 & 0.41 & 0.46 \\
C & $256^3$ &  5  &  466 & 0.6 & 2.2 & 0.9 & 2.5 & 0.39 & 0.46  \\
D & $256^3$ &  50 &  586 & 0.6 & 2.6 & 1.0 & 2.7 & 0.41 & 0.49  \\
E & $512^3$ &  1  &  426 & 0.4 & 2.0 & 1.0 & 2.0 & 0.45 & 0.49  \\
F & $512^3$ &  1  &  622 & 0.3 & 1.9 & 0.9 & 2.2 & 0.41 & 0.46  \\
G & $512^3$ &  10 &  675 & 0.5 & 2.2 & 0.8 & 2.5 & 0.39 & 0.44  \\
\hline
JHU & $1024^3$ & 1 & 1784 & & & 0.93 & 1.5 & \multicolumn{2}{|c|}{0.47} \\ 
\hline
\hline
\label{summary}
\end{tabular}
\end{table}

On the other hand, as can be seen from Fig.~\ref{sensPM}, the value of 
$\bar{\sigma}_{RM}$ in the saturated state is very
similar for all the runs we have considered. In particular
we again obtain $\bar{\sigma}_{RM}= 0.4-0.45$ (from method I) in the saturated
state, independent of $\Pm$, $\Rm$ and the resolution of the run.
Moreover, in all cases, as also discussed earlier, 
the start of an increase in $\bar{\sigma}_{RM}$ and $L_{int}$ 
from their values in the kinematic stage, is associated with the
onset of saturation due to Lorentz forces.

We have summarised the results of the RM computation for all
the runs in Table~\ref{summary}.  Looking at the last two columns 
of the table, we see that $\bar\sigma_{RM} \sim 0.4-0.5$ obtains 
in the saturated state
of the fluctuation dynamo almost universally.
It thus appears from our work that the effect of Lorentz force
is to order the field to a maximum scale which only depends on the
forcing scale, but is independent of the $\Pm$ and $\Rm$.
It would be important to do even higher $\Rm$ and $\Pm$ simulations
in the future to firm up these conclusions. 

\subsection{Introducing cutoffs}

The fluctuation dynamo generated fields are seen to be fairly intermittent,
especially if one looks at the high field regions \citep{HBD04,Schek04,BS05}.
An interesting question is to what extent the RM produced by such a field, 
arises in high field structures compared to the less intense
volume filling field regions. Addressing this issue could be important,
in case the high field regions are sensitive to $\Rm$ and $\Pm$. 
Thus it is useful to distinguish the RM contribution
from regions with differing field strengths.
Note that this can only be done from an actual realization
of the fluctuation dynamo generated field, and not by having
only the information about the magnetic power spectrum.

\begin{figure}
\epsfig{file=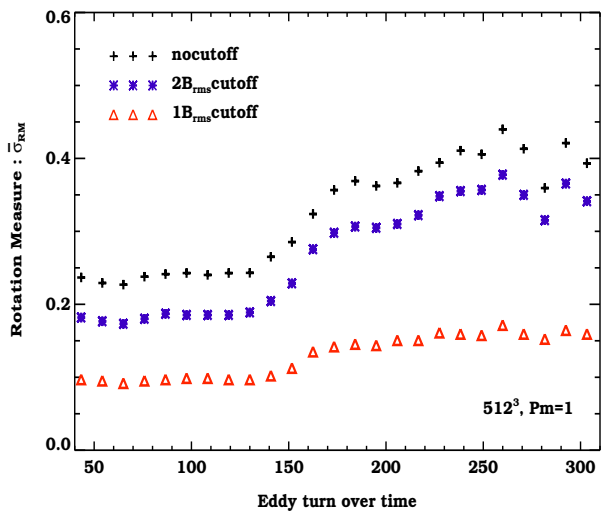, width=0.475\textwidth, height=0.3\textheight}
\caption{The time evolution of the normalised RM ($\bar\sigma_{RM}$)
for the $512^3$ run (F), determined excluding the regions with 
$\vert{\bf B}\vert > n B_{rms}$. The crosses correspond to not
imposing any cut-off, the stars show the result of excluding
$\vert{\bf B}\vert > 2B_{rms}$ regions, while the triangles show the
result of excluding regions with $\vert{\bf B}\vert > B_{rms}$.} 
\label{cutoff}
\end{figure}

\begin{figure}
\epsfig{file=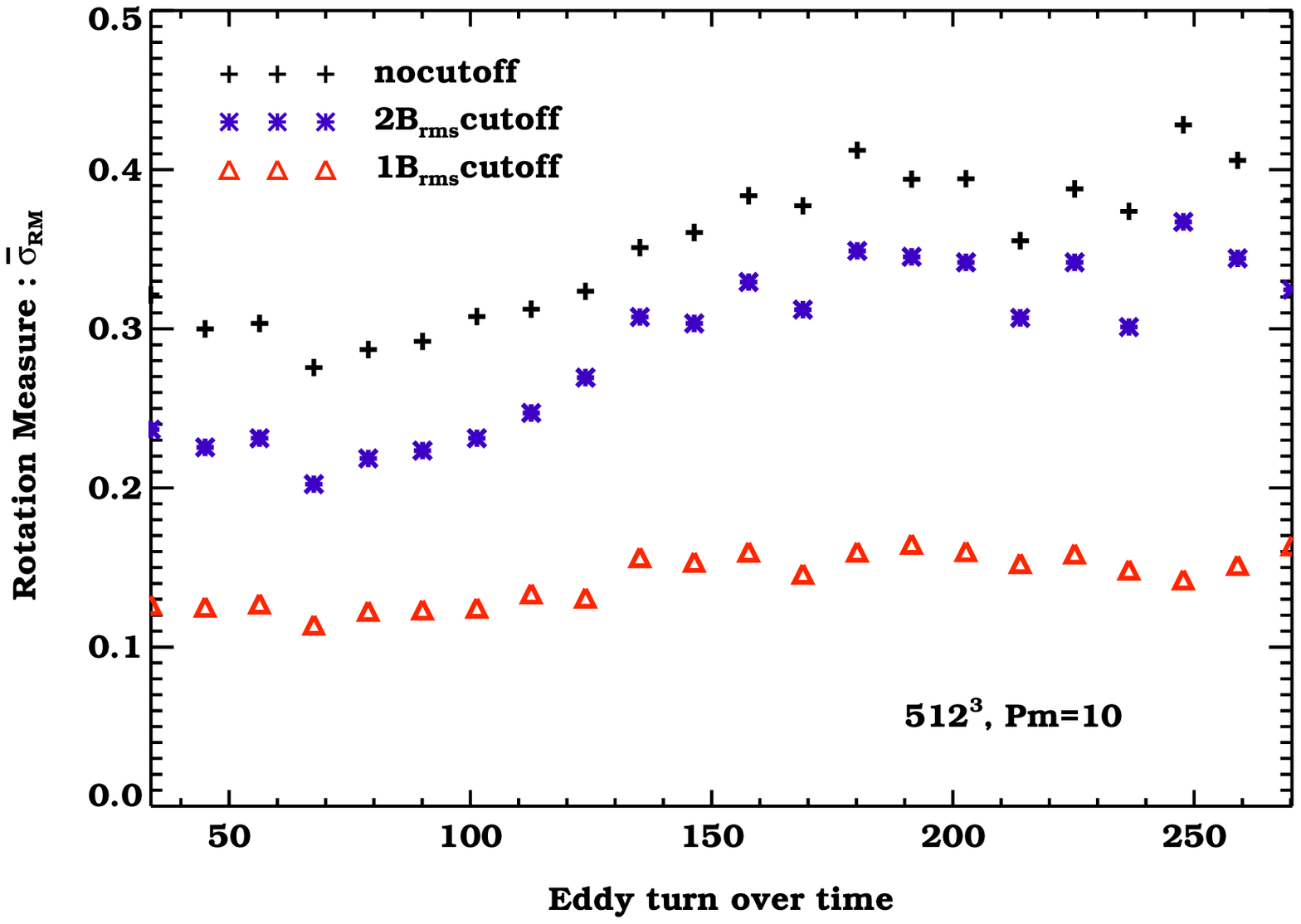, width=0.475\textwidth, height=0.3\textheight}
\caption{The time evolution of the normalised RM ($\bar\sigma_{RM}$)
for the $512^3$, $\Pm=10$ run (G), determined excluding the regions with 
$\vert{\bf B}\vert > n B_{rms}$. The crosses correspond to not
imposing any cut-off, the stars show the result of excluding
$\vert{\bf B}\vert > 2B_{rms}$ regions, while the triangles show the
result of excluding regions with $\vert{\bf B}\vert > B_{rms}$.} 
\label{cutoff10}
\end{figure}

We therefore calculate RM along each LOS, now 
leaving out regions where the field satisfies the constraint, 
$B^2= (B_x^2+B_y^2+B_z^2)>(nB_{rms})^2$, with $n=1$ and $2$. 
We repeat the same exercise of calculating the cumulative PDF,
$C(X)$, at each time, and finding the $\bar{\sigma}_{RM}$
after imposing the above constraint. Fig.~\ref{cutoff}
shows the results for the $\Pm=1$ run F,
while Fig.~\ref{cutoff10} shows corresponding results 
for the $\Pm=10$ run with $512^3$ resolution (run G).
The crosses in Fig.~\ref{cutoff} and Fig.~\ref{cutoff10} correspond to not
imposing any cut-off, the stars show the result of excluding
$B > 2B_{rms}$ regions, while the triangles show the
result of excluding regions with $B > B_{rms}$. 
The time evolution of $\bar{\sigma}_{RM}$ is shown right from
the kinematic stage up into saturation.

We find that $\bar{\sigma}_{RM}$ with a $B > 2B_{rms}$
cut-off has a similar time evolution to the case with no cutoff,
but with a reduced amplitude. 
From Fig.~\ref{cutoff}, for the $\Pm=1$ case, 
one finds that the regions with a field
strength larger than 2$B_{rms}$ contribute
only 15-20\% to the total RM. Thus the reduction in
the RMs calculated after one cuts off the 
$2B_{rms}$ fields is quite small. 
On the other hand, if one removes regions with
field strength larger than 1$B_{rms}$, then the RM decreases 
substantially, by a factor of 3 or so.
Moreover, the reduction in the RM with a cut-off is almost
the same, right from the kinematic stage to the saturated state.
This perhaps goes to show that
the fields generated by the $\Pm=1$ fluctuation dynamo 
grow in a self similar manner and the configurations do not 
change on the average substantially from the kinematic stage 
to saturation.


For the $\Pm=10$ case shown in Fig~\ref{cutoff10},
we find that the regions with a field
strength larger than 2$B_{rms}$ contribute
25\% to the total RM in the kinematic stage. However their
contribution reduces to about 15\% in the saturation stage, similar
to the $\Pm=1$ case.
Again then the RM decreases substantially, by a factor of 3
if one removes regions with field strength larger than 1$B_{rms}$.

These results thus show for both cases that the general `sea' of volume 
filling fluctuating fields contribute dominantly to the RM produced
by the fluctuation dynamo, rather than the high field regions,
right from the kinematic stage to the saturated state.

\section{Astrophysical applications}
\label{astro_app}

An important application of the above results is to galaxy
cluster plasma and their magnetic fields as inferred from radio
observations \citep{clarke_etal_01,carilli02,murgia04,govoni_feretti04,
vogt_ensslin,Govoni_etal2010,bonafede10,vacca10,bonafede11,
kuchar_ensslin11}. Note that 
the cluster magnetic fields will decay if not maintained 
by a turbulent dynamo of the nature considered in this work. 
As mentioned in the introduction, there is considerable evidence
from both observations and cosmological simulations that
cluster turbulence is nearly incompressible.
Thus our simulation results are directly applicable
in the context of explaining cluster magnetism.

Given the $\bar\sigma_{RM}$ obtained in our simulations,
one can estimate the expected dispersion of the RM, $\sigma_{RM}$,
in any given astrophysical system. This is given by
\begin{eqnarray}
\sigma_{RM} &=& 
\bar\sigma_{RM} \ K \ \overline{n}_e \ \frac{B_{rms}}{\sqrt{3}} \
l\sqrt{\frac{L}{l}} \ {\rm rad \ m}^{-2} \nonumber \\
&=& 180 \ {\rm rad} \ {\rm m}^{-2} \left(\frac{\bar\sigma_{RM}}{0.4}\right) 
\left(\frac{\overline{n}_e}{10^{-3} {\rm cm}^{-3}}\right)
\left(\frac{B_{rms}}{3 \mu {\rm G}}\right) \nonumber \\ 
&&\times \left(\frac{l}{100 {\rm kpc}}\right)^{1/2} 
\left(\frac{L}{1 {\rm Mpc}}\right)^{1/2}.
\label{dimfrm}
\end{eqnarray}
For obtaining a numerical estimate of $\sigma_{RM}$, 
we have adopted average values of various parameters appropriate for
a galaxy cluster (see below). 

In order to estimate $B_{rms}$ for the fluctuation dynamo generated
field, we note from Table~\ref{xxx} that on saturation the fluctation
dynamo generated field grows to a value of $B_{rms} \sim u_{rms}/2$, in
dimensionless code units. Therefore $B_{rms} \sim B_{eq}/2$ on saturation,
where $B_{eq}= \sqrt{4\pi\rho u_{rms}^2}$ is the 
field strength which is in equipartition
with the turbulent motions. This is given by 
\be
B_{eq} = 
6.1 \ \mu {\rm G} \left(\frac{n}{10^{-3} {\rm cm}^{-3}}\right)^{1/2}
\left(\frac{u_{rms}}{300 \ {\rm km \ s}^{-1}}\right),
\label{Beq}
\ee
where we have used $n=\rho/m_p$.

For a galaxy cluster, a typical value of $n=n_e \sim 10^{-3} cm^{-3}$,
$u_{rms} \sim 200-300$ km s$^{-1}$, 
$l\sim 100$ kpc and $L\sim 1$ Mpc (cf. 
SSH and references therein).
The eddy turn over time at the forcing scale is 
$\tau \sim l/u_{rms} \sim 3 \times 10^8$ yr. This $\tau$ is 
short compared
to the cluster age or the timescale for which mergers can sustain turbulence
(SSH, \cite{Ryu_etal2012}). Therefore  
magnetic fields are likely to be amplified to the saturation value 
by the fluctuation dynamo. 
One may then expect 
$B_{eq} \sim 4-6 \ \mu {\rm G}$, and a fluctuation dynamo generated
field with $B_{rms} \sim 2-3 \ \mu {\rm G}$ in a galaxy cluster.  
Then Eq.~\ref{dimfrm} predicts an rms value of RM,  
$\sigma_{RM} \sim 120-180$ rad m$^{-2}$ in galaxy clusters.
The data presented by 
\cite{clarke_etal_01} show a typical scatter in the RM values
for LOS through galaxy clusters of 
$\sim 100 \ {\rm rad \ m}^{-2}$. 
Therefore one sees that the average value of
$\bar\sigma_{RM}$ obtained from our simulation of
the fluctuation dynamo, is sufficiently large to account for
the RM measured in galaxy clusters. 

The above average estimates can be generalised to 
the situation where the $n_e$ and $B_{rms}$ depend on the
cluster radius. We have considered this in detail
in Appendix~\ref{app}. 
Assuming that the correlation scale
of the turbulence is small compared to the cluster scales,
it turns out that this simply involves replacing 
$\bar{n}^2_e B^2_{rms} L$ 
in Eq.~\ref{dimfrm}, when squared, by the integral I,
given by,
\begin{equation}
I({\bf r}_\perp)=\int n^2_e({\bf r}_\bot,Z) B^2_{rms}({\bf r}_\bot,Z) dZ
\end{equation}
where the LOS is parallel to the $Z$-direction and
${\bf r}_\bot$ is the perpendicular displacement from 
the center of the cluster.
Then the $\sigma_{RM}$ for such a model is given by Eq.~\ref{rmdisp3},
which after using $L_L=3L_{int}/8$ and $L_{int}=4 \ \bar\sigma^2_{RM}l/3$ 
from Eq.~\ref{cr_frm}, can be written as,
\begin{equation}
\sigma^2_{RM} =
\bar\sigma^2_{RM} \ \frac{K^2 \ l}{3} \ I \ {\rm rad} \ {\rm m}^{-2} 
\end{equation}
Using the standard $\beta$-model for the density profile of a cluster, 
$n_e \propto (1+r^2/r_c^2)^{-3\beta/2} $ ($r_c$ is the cluster core 
radius) and assuming, 
$B_{rms} \propto n_e^{\gamma}$, we can evaluate the integral, I,
exactly (see Appendix~\ref{app}). We get,
\begin{equation}
\sigma_{RM}(r_\bot)
= \sigma_{RM}(0)
\left(1+\frac{r^2_{\bot}}{r_c^2}\right)^{-\frac{(6\beta(\gamma +1)-1)}{4}}, 
\label{comprehensive}
\end{equation}
where
\begin{equation}
\sigma_{RM}(0) = \bar\sigma_{RM}
\frac{\pi^{1/4}}{\sqrt{6}} \ K  n_0  B_0 \sqrt{r_c \ l} \
\sqrt{{\frac{\Gamma(3\beta(\gamma +1)-0.5)}{\Gamma(3\beta(\gamma +1))}}}
\end{equation}
and $n_0$, $B_0$ are the central density
and rms magnetic field strength respectively.

As an example, consider Coma cluster, where we adopt 
from \citet{bonafede10},
$n_0=3.44\times10^{-3} {\rm cm}^{-3}$, $\beta=0.75$, $r_c=291$ kpc 
and a constrained 
$B_0=3.9 \ \mu {\rm G}$ for $\gamma=0.4$.
For these values, and assuming $l=100$ kpc, 
the $\sigma_{RM}$ for a source seen through the 
Coma cluster at an impact parameter distance,
$r_\perp = 50\ {\rm kpc}$, is estimated to be, 
$\sigma_{RM} \sim 310 \ (\bar\sigma_{RM}/0.4) \ {\rm rad} \ {\rm m}^{-2}$,
which is close to the value $\sigma_{RM,obs} = 303 \ {\rm rad} \ {\rm m}^{-2}$ 
observed, as quoted in \citet{bonafede10}.
Note that this crucially depends on the normalised
$\bar\sigma_{RM}$ or the magnetic field
correlation scale as determined from the fluctuation dynamo
simulations being large enough; which we have shown here is indeed
the case. Importantly, for the case of Coma, \cite{bonafede10} also
note the magnetic field as determined from RM measurements 
averaged over a Mpc$^3$ volume is compatible with 
equipartition estimates obtained from modelling the Coma radio halo.
Thus for the case of Coma, a picture whereby magnetic fields
are amplified by a fluctuation dynamo driven by incompressible
turbulence, seems consistent with radio observations.

The fluctuation dynamo can also lead to magnetic field generation
in intergalactic filaments at the present epoch. 
Here, vorticity and turbulence are generated in shocks resulting 
from large scale structure formation
\citep{Ryu_etal2008,Iapichino_etal2011,Ryu_etal2012}. 
Combining the estimated levels of the resulting turbulence 
with a model of magnetic field generation by the fluctuation dynamos, 
\citet{Ryu_etal2008,CR09} estimate a magnetic field 
of tens of nG in these filaments and their RM contribution to be
$\sim 1$ rad m$^{-2}$. A crucial question is again, how coherent
is the dynamo generated field, which has been the focus of our work.

Consider now the case of the interstellar medium of a gas rich 
disk galaxy, possibly at high redshift. 
In galaxies supernovae typically drive the turbulence, and even though
the forcing may be largely compressible and at high mach numbers, the
intersection of shocks and shock propagation through the inhomogeneous
ISM leads to vorticity generation. The resulting vortical
turbulent motions can again drive a fluctuation dynamo and
amplify magnetic fields 
\citep{korpi_etal1999,haugen_axel_mee04,avillez05,balsara05,gressel,
wu_etal09,federrarth11,gent12}.
Note that if a gas mass
$M\sim 10^{10} M_\odot$ is distributed in a disk of radius
$r=10$ kpc and thickness $2h=1$ kpc, the average density
$n\sim 1.4$ cm$^{-3}$. For example, the total stellar mass
in our Galaxy is $\sim 6\times 10^{10} M_\odot$ \citep{Binney_Tremaine}
and the gas mass could be about $10\%$ of this value. For such a galaxy
one would get $n\sim 0.84$ cm$^{-3}$. 
One may have a higher gas mass fraction for a high redshift disk galaxy. 
We adopt $n=1$ cm$^{-3}$ to estimate $B_{eq}$.
A caveat is that the turbulence in the ISM is expected to be
transonic. In principle, for such turbulence, density fluctuations 
correlated with the field could affect the RM estimates. 
However, several simulations of supernovae driven turbulence do not find 
a very strong correlation between magnetic field and density 
\citep{avillez05,wu_etal09} and so, we expect our RM estimates 
to be reasonably indicative.

Let us adopt a
typical vortical turbulent velocity, $u_{rms} \sim 10$ km s$^{-1}$
of order the sound speed in the warm ionised interstellar medium (ISM), 
and turbulent forcing scale $l\sim 100$ pc \citep{korpi_etal1999,shuk04}.
Then the eddy turn over time $\tau \sim 10^{7}$ yr, is much less
that the age of disk galaxies, even at high redshifts. Thus
one expects the fluctuation dynamo to grow the magnetic
field to saturation even for weak seed fields. 
This then gives $B_{eq} \sim 6.5 \ \mu {\rm G}$
and $B_{rms}$ could be a fraction $f$ of this value.
For a line of sight
of length $L=1$ kpc through the disk thickness, 
and $f\sim 1/2$, 
we get from
Eq.~\ref{dimfrm}, $\sigma_{RM} \sim 180$ rad m$^{-2}$. 
Thus, again, significant Faraday rotation 
is expected if a line of sight from a background radio source
passes through a gas rich disk galaxy, even if the fields produced
in such a disk is purely through fluctuation dynamo action. 
In other words, observations
of significant RM at high redshift need not require
the canonical mean field helical dynamo to have generated
large scale coherent fields.
We emphasize that these results are only indicative and one
requires much more work on fluctuation dynamo action
in SNe driven turbulence to substantiate the above conclusions.
It is also perhaps worth noting that significant magnetic
field generation could occur even before forming the
disk, due to the fluctuation dynamo action in the turbulent halo gas
as the galaxy forms,  
in manner similar to what we have discussed in cluster plasma 
\citep{kulsrud_etal97,arshakian09,Schleicher10,schober12,
beck_etal12,sur_etal12}.
Again whether this produces coherent enough fields is the
crucial issue, one which we have answered in the affirmative 
in our work here.

Note that only some of the MgII absorption systems 
probed by \citet{Bernet08} are likely to arise in
lines of sight through a galaxy disk. It is believed that
many of these sight-lines could also be sampling the gaseous
halo around a massive galaxy (cf. the review by \citet{church2005})
or even an associated smaller dwarf 
galaxy only detected by 
spectral stacking \citep{noter_anand10}. 
The halo gas is likely to be hot, either accreted during the formation of
the galaxy, or transported out of the disk in a 
supernovae driven wind or fountain flow 
(see for example \citet{nestor11,bouche12}). This halo medium
needs to contain not only the hot gas but also entrained 
magnetised cool gas which produces both the
MgII absorption, and the excess RM seen by \citet{Bernet08}.
Alternatively, the much cooler and magnetised material could be driven out
by the pressure of cosmic rays as in the wind 
models of \citet{samui10}.
We are assuming here that the magnetisation of the gas takes place
in the disk by say the fluctuation dynamo, and then this gas is 
ejected in the wind, along with the metals. Alternatively if the
hot wind is turbulent, then the fluctuation dynamo can operate
in the wind itself.

We model such MgII systems
by assuming say that the
line of sight through the gaseous halo
passes through $M$ magnetised (and MgII rich)
`clouds' each of scale $l$, electron density $\bar{n}_e$ 
and with an average field
of strength $B_{0}$, which is again randomly oriented
between cloud to cloud. Then an analysis identical to that
which gave Eq.~\ref{frmo} can be applied. The resulting 
rms value of RM, $\sigma_{RM}$, through such a line of sight
will be given by
\begin{eqnarray}
\sigma_{RM} 
&=&  K\bar{n}_e \frac{B_{0}}{\sqrt{3}} \ l\sqrt{M}
= K N_e \frac{B_{0}}{\sqrt{3M}} \nonumber \\
&=& \frac{160}{\sqrt{M}} \ {\rm rad} \ {\rm m}^{-2} 
\left(\frac{N_e}{ 10^{20} {\rm cm}^{-2}}\right)
\left(\frac{B_{0}}{10 \mu {\rm G}}\right). 
\label{frmmg2}
\end{eqnarray}
Here, $N_e = (\bar{n}_e l)M$ is the total electron column density through
the $M$ magnetised clouds. The magnetic field in a cloud, which is
denser than the average ISM, could be larger than the $B_{rms}$
estimated from fluctuation dynamo action in the average density ISM.
For example, if the cloud density is 10 times larger, (like
say a compact HII region in the ISM) then
assuming flux freezing, $B_0 \sim 10^{2/3} B_{rms} \sim 15 \ \mu {\rm G}$. 
\citet{Bernet08} estimate 
$N_e \sim 10^{20}$ cm$^{-2}$ for their MgII systems, while the
multiplicity of components $M$ will vary from system to system. 
Then adopting $B_0\sim 5-15\ \mu {\rm G}$, we see from
Eq.~\ref{frmmg2} that $\sigma_{RM} \sim (80-230)/\sqrt{M}$ rad m$^{-2}$. 
This will be within about the 1$\sigma$ value inferred by \cite{Bernet08},
who find $\sigma_{RM} \sim 140^{+80}_{-50}$ rad m$^{-2}$, for $M< 7$.
Thus the level of RM excess detected in the MgII systems, seems 
marginally consistent with theoretical expectations.

\section{Discussions and conclusions}

There is considerable evidence for the presence of coherent
magnetic fields in various astrophysical systems, from galaxies to 
galaxy clusters.
Much of this evidence comes from measurements of Faraday rotation.
These systems are also generically 
turbulent and would therefore host what are referred to as 
fluctuation or small scale dynamos. 
Such a dynamo amplifies magnetic fields on the
fast eddy turn over timescales. However the generated fields
are believed to be intermittent.
We have considered whether the fluctuation dynamo generated fields
can nevertheless lead to a sufficient degree of Faraday rotation so as to
explain the observations. This is especially important for
systems which are either too young, or do not have the required 
conditions, for significant amplification of the field 
by a large-scale dynamo. 

For this purpose, we have run a suite of fluctuation dynamo
simulations in periodic boxes, with resolutions of up to $512^3$,
and for a range of $\Rm$ and $\Pm$.
We can then directly calculate the 
time evolution of the
Faraday rotation measure (RM)
predicted by these simulations
from the kinematic to the saturated state of the fluctuation dynamo.
We have used 3 different methods
for this purpose. 
In the first method (I) we shoot $3N^2$ lines of sight 
through the simulation box,
form cumulative PDFs of the measured RMs and estimate its 
normalised dispersion $\bar{\sigma}_{RM}$, as the 1$\sigma$ range 
contain the
central $68.2$\% of RM values. We have also directly calculated the 
standard deviation of the measured RMs (method II). 
Finally, in method III, we have
estimated the RM dispersion using the magnetic energy spectrum
under the assumption of statistical isotropy. As shown in
Fig.~\ref{RMc}, all 3 methods give very similar results,
with method II giving $\sim 10\%-15\%$ larger  $\bar{\sigma}_{RM}$,
and thus for all subsequent results of $\bar{\sigma}_{RM}$, 
we mostly
use method I.

On analysing the suite of fluctuation dynamo simulations, we show that
the value of $\bar{\sigma}_{RM}$ after the dynamo saturates
is very similar for all the runs. In particular
$\bar{\sigma}_{RM}= 0.4-0.5$ in the saturated
state is independent of $\Pm$, $\Rm$ and the resolution of the run
(see Fig. ~\ref{sensRM}, \ref{sensPM} and the last two columns of 
Table~\ref{summary}).
In addition, $\bar{\sigma}_{RM}$ of this order also obtains 
for an independent higher resolution ($1024^3$) 
and higher Reynolds number 
simulation
of fluctuation dynamo from the JHU database (last row of 
Table~\ref{summary}).
This is a fairly large value for an intermittent random field;
as it is of order 40\%-50\%, of that expected in a model
where $B_{rms}$ strength fields volume fill each turbulent
cell, but are randomly oriented from one cell to another.

We also find that the regions with a field
strength larger than 2$B_{rms}$ contribute
only 15-20\% to the total RM. Thus the reduction in
the RMs calculated after one cuts off the 
$2B_{rms}$ fields is quite small. 
On the other hand, if one removes regions with
field strength larger than 1$B_{rms}$, then the RM decreases 
substantially, by a factor of 3 or so. These
numbers obtain for both the $512^3$ $\Pm=1$ and $\Pm=10$ runs.

The fact that cutting out the large field regions does
not significantly reduce the RM resulting from fluctuation 
dynamo generated fields, suggests the following picture.  
It shows that it is the general `sea' of volume filling 
fluctuating fields that contribute dominantly to the RM produced
by the fluctuation dynamo, rather than the high field regions,
right from the kinematic stage to the saturated state.

Moreover, in all cases, we find that $\bar{\sigma}_{RM}$ and $L_{int}$
begin to increase from their value in the kinematic stage, at the onset 
of saturation, when the influence of Lorentz forces becomes important.
Therefore, the effect of Lorentz forces due to the fluctuation dynamo
generated field, is to order the field to larger and larger scale 
up to almost a universal maximum value. This maximum value seems to only
depend on the forcing scale, and importantly is independent of 
the $\Pm$ and $\Rm$ to the extent we have tested.
It would be important to do even higher $\Rm$ and $\Pm$ simulations
in the future to firm up these conclusions. 

Note that from Eq.~\ref{cr_frm}, the dispersion in the normalised 
RM ($\bar\sigma_{RM}$), is related to the integral scale of the
magnetic field, $L_{int}$, as defined in Eq.~\ref{lint}.
Thus, the discussions above bring to fore, also the evolution of 
the magnetic integral
scale in the simulations of fluctuation dynamos. 
In the kinematic stage, $L_{int}$ does depend on $\Rm$ and $\Pm$;
we find that lower the $\Rm=\Rey$ or higher the $\Pm$, 
larger is the magnetic integral scale in kinematic stage. 
This can be understood by studying the $\Rm(k)$ spectra
and the eddy turn over rate spectra (see Fig~\ref{eddy:spectra:kin}).
We find that the first eddies which amplify the field efficiently 
are larger for runs with higher $\Pm$ or lower $\Rm=\Rey$, as they have 
a larger turn over rate and 
also their corresponding $\Rm(k)$ is supercritical.
This then leads also to a correspondingly larger $L_{int}$
for cases with lower $\Rm=\Rey$ or higher $\Pm$.

The situation when the dynamo saturates is quite different.
Although the manner in which fluctuation dynamos saturate 
is not the main focus of our work, our results
on the evolution of the magnetic integral scale, $L_{int}$, 
point to some interesting features of saturation. 
Firstly, in the $512^3$, $\Pm=1$ case (run F), $L_{int}$ 
increases from about $0.3$ in the kinematic stage to $\sim 0.9$ 
in saturated stage, or by a factor of about 3.
Such an increase, if not of the same magnitude,
can be seen in all the runs, and begins always, when the
Lorentz forces become important.
In contrast the integral scale of the velocity field does not
change appreciably in any of the runs. 
On saturation, the magnetic integral scale 
$L_{int}$ is only a factor of $2-3$ smaller than velocity integral scale 
$L_{int}^V$ (defined in an identical manner).
Also, the integral scales of the saturated magnetic field are 
very similar, with $L_{int} \sim 1$, for all our runs 
though they have different $\Rm$ and $\Pm$.
These features suggest that  
the integral scale of the magnetic field in the saturated state
does not depend on the microscopic resistivity or viscosity.
It therefore appears that
Lorentz forces can indeed order the magnetic 
field and increase its coherence scale 
to be a modest fraction ($\sim 1/2-1/3$) of the velocity coherence 
scale as the fluctuation dynamo saturates. 

The dispersion of the normalised RM obtained here $\sim 0.4-0.5$ 
implies a dimensional $\sigma_{RM} \sim 180$ rad m$^{-2}$, for
parameters appropriate for galaxy clusters. This is
sufficiently large to account for the observed Faraday rotation
seen in these systems. One also obtains a similar estimate for
lines of sight through a disk galaxy.
The fluctuation dynamo will generate the first fields
in any turbulent system like a young galaxy. Our result that the
generated field is fairly coherent and can lead to significant
RM, even in the absence of a mean field generation, will be of
interest when one detects RM from higher and higher redshift galaxies. 
The present detection of excess RM from MgII systems, is
marginally consistent with theoretical expectations.

Note that our work is complementary to those which 
simulate large scale structure formation including the formation
of massive galaxy clusters and the resulting magnetic field 
generation \citep{Dolag2006,Xu_etal2009,Ryu_etal2012,Xu_etal2012}.
The cosmological simulations typically have a modest resolution of
a turbulent eddy, as they have to also accommodate scales of the
order a cluster radius and larger. On the other hand,
we have driven the turbulence at a scale
comparable to the box scale, so as to resolve the small scale structure
of the magnetic field as well as possible within a turbulent cell.
Both types of simulations will be useful to get a complete
picture of the magnetic field generation in say galaxy clusters.

Astrophysical systems typically have 
much higher Reynolds numbers than any simulation would be 
able to achieve in the near future. However, some of the basic 
features of fluctuation dynamos are expected to be stable
to the increase in Reynolds numbers. There are
simulations of fluctuation dynamos with $\Pm=1$, at higher 
resolutions (implying higher Reynolds numbers) than what we
have presented here \citep{HBD03,HBD04,Li_etal2008,Perlman_etal,
jones11}, which show qualitatively very similar kinetic and 
magnetic spectra. Infact, we find that the $\bar\sigma_{RM}$ 
from the JHU $1024^3$ simulation (table \ref{summary}) match our results. 
It appears that for $\Pm=1$, resolution of current simulations 
is sufficient to get converging results. Systems with large $\Pm$ 
and large $\Rey$ are more difficult to simulate and would require
improved computing resources. Nevertheless, to the extent we have
explored larger $\Pm$ case, the value of $\bar\sigma_{RM}$ and hence
the field coherence  
appears consistent with the $\Pm=1$ case.
  
We have concentrated in the present paper on the RM signals
from the fluctuation dynamo generated fields. It will also
be of interest to study other observables, like the
synchrotron emissivity and polarization signals.
As the synchrotron emissivity depends nonlinearly 
on the field strength, these signals would be more sensitive
to the more intense and rarer structures, compared to
the RM signal. It will also be of great interest to 
explore the results obtained here with higher resolution simulations,
and also obtain an improved understanding of how fluctuation dynamos
saturate, which remains a challenge.

\section*{Acknowledgments}

PB thanks Sharanya Sur for initial help with the Pencil code.
We thank Greg Eyink for alerting us to the JHU database. We also
thank Greg Eyink, Axel Brandenburg and R. Srianand for very useful discussions. 
We acknowledge the use of the HPC facility at IUCAA.
KS acknowledges partial support from NSF
Grant PHY-0903797 while at the University of Rochester.
KS thanks Eric Blackman and Greg Eyink for warm hospitality
at Rochester and Baltimore during his visit there.
We thank an anonymous referee for comments which
have led to improvements to our paper.
\bibliographystyle{mn2e}
\bibliography{frmrefs}
\appendix

\section{General analytical description of RM dispersion}
\label{app}

The Faraday rotation measure for a LOS, `L' from the source
to the observer, parallel to z-axis
and perpendicularly displaced by ${\bf r}_{\bot}$ from the center of the cluster,
is given by,
\begin{equation}
{\rm RM} =  K \int_L n_e({\bf r}_{\bot}, z) \ B_z({\bf r}_{\bot}, z) dz,
\label{FRMcluster}
\end{equation}
Here $B_z$ is the component of the magnetic field in the z-direction.
Then, the dispersion in RM along the LOS is given by,
\begin{eqnarray*}
\langle \left({\rm RM}\right)^2\rangle &=&  
K^2 \int \int n_e({\bf r}_{\bot}, z) 
\ n_e({\bf r}_{\bot}, z') \\ \nonumber
 && \ \ \ \ \ \ \ \ \ \ \ \ \ \ \ \ \ \overline{B_z({\bf r}_{\bot}, z) \ 
B_z({\bf r}_{\bot}, z^{'})} 
\ dz \ dz', 
\label{rmdisp}
\end{eqnarray*}
where we have assumed that the density and magnetic field are 
not correlated.
We define the relative co-ordinate $r_z=z'-z$ and the mean $Z=(z+z')/2$.
For a random weakly inhomogeneous and isotropic magnetic field, we write 
the magnetic correlator as
$\overline{B_i({\bf x}) \ B_j({\bf y})}=M_{ij}({\bf r};{\bf R})$, 
where ${\bf r}=({\bf x} - {\bf y})$ and ${\bf R} = ({\bf x} + {\bf y})/2$.
This correlator is supposed to vary rapidly with ${\bf r}$ but
slowly with ${\bf R}$. We will also assume that the correlation scale
of the field is small compared to the variation scale of the density.
Then the RM dispersion can be written as
\begin{equation}
\langle \left({\rm RM}\right)^2\rangle
= K^2 \int_{-\infty}^{\infty} dZ \  
n_e^2({\bf r}_{\bot}, Z) 
\int_{-\infty}^{\infty} dr_z \
M_{zz} (r_z;{\bf r}_{\bot}, Z), 
\label{rmdisp2}
\end{equation}
${\bf R} \equiv ({\bf r}_\bot,Z)$.
Note that $M_{zz}(\vert r_z\vert)$ is exactly the longitudinal
correlation function $M_L(r)$ (cf. S99). 
We define the longitudinal integral scale in the standard fashion as, 
\[
L_L=\frac{\int_0^\infty M_{zz}(\vert r_z\vert;{\bf r}_{\bot}, Z ) \ dr_z}
{M_{zz}(0; {\bf r}_{\bot}, Z)}.
\]
Moreover, as $M_L(0; {\bf r}_{\bot}, Z) = B_{rms}^2({\bf r}_{\bot}, Z)/3$,
we get
\begin{equation}
\langle \left({\rm RM}\right)^2\rangle =
\frac{2K^2}{3}  \int_{-\infty}^{\infty} L_L \ n_e^2({\bf r}_{\bot}, Z) B_{rms}^2({\bf r}_{\bot}, Z) \ dZ,
\label{rmdisp3}
\end{equation}
where the factor 2 arises because the integral over $r_z$ in Eq.~\ref{rmdisp2}, 
over the intervals $(0,\infty)$ and $(-\infty,0)$ contribute equally to $L_L$.
%

For a density and rms magnetic field of uniform strength across the LOS
with length $L$,
the Eq.~\ref{rmdisp3} reduces to,
\begin{equation}
\left\langle ({\rm RM})^2\rangle\right)^{1/2} = \sigma_{RM}= 
 K n_e B_{rms}
\frac{\sqrt{L_{int}L}}{2}
\end{equation}
where we have substituted $L_L=3L_{int}/8$. 
This equation is identical to Eq. (10) of 
\cite{CR09}, which they derived from a Fourier space
analysis, and it leads to the Eq.~\ref{cr_frm} that
we have used in calculation of RM as per method III.

Consider now the more general case when the density and rms magnetic field
varies with the cluster radius 
We use the standard $\beta$-model for the electron number
density profile, $n_e \propto (1+r^2/r_c^2)^{-3\beta/2}$ 
($r_c$ is the cluster core radius) 
and also assume $B_{rms} \propto n_e^{\gamma}$. Then 
the dispersion in RM becomes,
\begin{equation}
\langle ({\rm RM})^2\rangle = \frac{2L_L}{3} \ K^2 \ n^2_0 \ B^2_0  
\int_{-\infty}^{\infty} \frac{dZ}{\left(1+\frac{r^2_{\bot}+Z^2}{r^2_c}\right)^{3\beta(\gamma +1)}},
\end{equation}
where $r_\bot=\vert{\bf r}_\bot\vert$. Also $n_0$ and 
$B_0$ are the central density
and rms magnetic field strength respectively.
The integral over $Z$ can be evaluated exactly to give
(Eq. 4 of 3.241 in \cite{grad_book}).
\begin{equation}
\sigma_{RM}(r_\bot)
= \sigma_{RM}(0)
\left(1+\frac{r^2_{\bot}}{r_c^2}\right)^{-\frac{(6\beta(\gamma +1)-1)}{4}}, 
\label{comprehensive2}
\end{equation}
where
\begin{equation}
\sigma_{RM}(0) =
\frac{\pi^{1/4}}{\sqrt{3}} \ K  n_0  B_0 \sqrt{r_c \ L_L} \
\sqrt{{\frac{\Gamma(3\beta(\gamma +1)-0.5)}{\Gamma(3\beta(\gamma +1))}}}
\end{equation}
The same expression with the longitudinal correlation
scale $L_L$ replaced by the cell size $l$ is given by
\cite{Felten}, for the case $\gamma=0$.
  
\label{lastpage}

\end{document}